\documentclass{article}
\usepackage{amsmath}
\def\p{\partial}
\def\s{\sigma}

\def\g{\gamma}

\def\d{\delta}
\def\de{\delta}
\def\D{\Delta}

\def\ld{\lambda}
\def\Ld{\Lambda}
\def\L{\Lambda}
\def\ep{\epsilon}
\def\e{\eta}

\def\om{\omega}

\def\rh{\rho}
\def\z{\zeta}

\def\b{\beta}

\def\a{\alpha}

\def\pdellx'{\frac{\partial}{\partial x'}}
\def\pdellw'{\frac{\partial}{\partial w'}}
\newcommand{\be}{\begin{equation}}
\newcommand{\ee}{\end{equation}}
\def\bed{\begin{displaymath}}
\def\eed{\end{displaymath}}
\def\bea{\begin{eqnarray}} 
\def\eea{\end{eqncrray}}
\def\[{$$}
\def\]{$$}
\begin{document}
\title{ Exact Space-Time Gauge Symmetry of Gravity, Its Couplings and Approximate Internal Symmetries in a Total-Unified Model}%\bigskip
\author{ Jong-Ping Hsu\footnote{e-mail: jhsu@umassd.edu}\\
Department of Physics,\\
 University of Massachusetts Dartmouth,\\
 North Dartmouth, MA 02747-2300, USA}
\maketitle
{\small Gravitational field is the manifestation of space-time translational ($T_4$) gauge symmetry, which enables gravitational interaction to be unified with the strong and the electroweak  interactions.  Such a total-unified model is based on a generalized Yang-Mills framework in flat space-time.  Following the idea of Glashow-Salam-Ward-Weinberg, we gauge the groups $T_4 \times (SU_3)_{color} \times SU_2 \times U_1\times U_{1b}$ on equal-footing, so that we have the total-unified gauge covariant derivative ${\bf \d}_{\mu} = \p_{\mu} - ig\phi_{\mu}^{\nu} p_{\nu}+ig_{s}{G_{\mu}^{ a}}( \ld^a/2) 
+if{W_{\mu}^{ k}}{ t^k} + if' U_{\mu}t_{o} + ig_{b}B_{\mu}$. The generators of the external $T_4$ group have the representation $p_{\mu}=i\p_{\mu}$, which differs from other generators of all internal groups, which have constant matrix representations.  Consequently, the total-unified model leads to the following new results: (a)  All internal $(SU_3)_{color}, SU_2, U_1$ and baryonic $U_{1b}$ gauge symmetries have extremely small violations due to the gravitational interaction.  (b)  The $T_4$ gauge symmetry remains exact and dictates the universal coupling of gravitons. (c)  Such a gravitational violation of internal gauge symmetries leads to modified eikonal and Hamilton-Jacobi type equations, which are obtained in the geometric-optics limit and involve effective Riemann metric tensors.  (d)  The  rules for Feynman diagrams involving new couplings of photon-graviton, gluon-graviton and quark-gaviton are obtained.}
\bigskip	

\section{Introduciton}

In previous papers, we have discussed electroweak and strong sectors of a unified model involving Yang-Mills gravity with space-time translational gauge group $T_4$ based on flat space-time.\cite{1,2}  We show that an experimentally consistent model, which unifies Yang-Mills gravity and other forces in nature, can be formulated on the basis of a gauge invariant action under local space-time translations.  For  the motion of classical  objects,  eikonal (or Hamilton-Jacobi type) equations are derived from $T_4$ gauge invariant  wave equations of bosons and fermions in the geometric-optics limit, and they contain `effective Riemannian metric tensors.'   The emergence of  `effective curved space-time with Riemannian metric tensors' in (and only in) the geometric-optics limit of  wave equations is essential for Yang-Mills gravity to be formulated in flat space-time and  consistent with experiments.  This approach brings gravity back to the arena of gauge field theory and quantum mechanics in flat space-time.\cite{3,4}  

In this paper, we step back and look at the big picture of all possible interaction forces in nature and explore the most general implications of a total-unified model based on a generalized Yang-Mills framework, which can accommodate both internal and external gauge groups in flat space-time.  We know that the electromagnetic and weak interactions have been unified based on a gauge invariant action with the internal gauge groups $ SU_2 \times  U_1$.\cite{5}  And the strong interaction can also be described a gauge field theory, i.e., by quantum chromodynamics with internal color $SU_3$ groups.  Apart from these well-known forces, there may be other forces of gauge fields related to, say, accelerated cosmic expansion\cite{6,7} and quark confinement.\cite{8}  For example, the cosmic repulsive force due to a gauge field works if and only if there is not much anti-matter in the observable universe.   Moreover, such a force should be much weaker than gravity within the solar system and our galaxy, but its repulsive force between two galaxies should become stronger than the gravitational force. We have discussed a modified Lee-Yang  $U_{1b}$ baryonic gauge field, which satisfies a gauge invariant fourth-order differential equation and has a linear static potential.\cite{8}  Such a modified Lee-Yang model predicts a constant acceleration for the expansion of the universe, provided the observable universe are dominated by baryon matter.   Thus, it appears to suggest that all known interactions can be understood on the basis of gauge symmetries and, hence, can be unified, similar to the unification of the electromagnetic and weak interactions.  

At the first glance of the total-unified model by gauging the groups,
$T_4 \times (SU_3)_{color} \times SU_2 .....$, one may get the impression that, in the total covariant derivative, various gauge fields with their coupling constants and group generators are simply juxtaposed, without any true integration.  
However, such a total-unified model without any true integration is probably all what one can do in the physical world as we know it.  The reasons are as follows: (i) Mathematically, it appears to be very unlikely that one can further unified these external and internal gauge symmetries in one single Lie group.  (ii)  The usual asymptotic freedom and running coupling constants in non-Abelian gauge theories are no longer true in the total-unified model, which is not renormalizable by power counting because the gravitational interaction has the dimensional coupling constant $g$.  The asymptotic freedom of non-Abelian gauge theories hold only when one ignores gravity.  But no physical particle in nature can escape from the gravitational interaction.  Therefore, it is very unlikely that all different interactions can be unified at extremely high energies with one single coupling constant.   Nevertheless, the total-unified model involving Yang-Mills gravity\cite{9,10} is interesting because it leads to a new understanding of physics that gauge symmetries dictate all particle interactions\cite{11}  and can make new predictions to be tested experimentally.

\section{Total unification of all interactions and gauge curvatures }

Based on previous discussions of Yang-Mills gravity,\cite{9} a total-unified group $G_{tu}$ in flat space-time, consistent with experimentally established conservation laws, appears to be\cite{1,2}
\be
G_{tu}=T_4  \times [SU_3]_{color} \times \{SU_2 \times U_1\} \times U_{1b},
\ee
%%%%%%%%C2%16.27%%%1
where, if one wishes, the leptonic conservation law can also be included in (1) with an additional leptonic $U_1$ group.  The mathematical structure of the leptonic gauge field is identical to that of the baryonic gauge field, so we shall not consider it here.
 In the total-unified model, the total gauge covariant derivative ${\d}_{\mu} $ in  general frames (inertial and non-inertial) of reference\cite{4,9,12}  (with the Poincar\'e metric tensor $P_{\mu\nu}$ and the associated covariant derivative $D_{\mu}$) is assumed to take the form (c=$\hbar$=1)
\be
\d_{\mu} = D_{\mu} + g\phi_{\mu}^{\nu}D_{\nu}
%%%%%%%%%%C3%16.28
+ig_{s}{G_{\mu}^{ a}}\frac{{ \ld^a}}{2} 
+if{W_{\mu}^{ m}}{ t^m} + if' U_{\mu}t_{o}  +\frac{ig_{b}}{3}B_{\mu},  
\ee
%%%28%%%%%%%%%%2
where $t^m$,  $m=1,2,3$, are $SU_2$ generators in general and $t_o$ (weak hypercharge) is the $U(1)$ generator.\cite{5}  In the limit of zero acceleration, all non-inertial frames reduce to inertial frames,\footnote{The Poincar\'e metric tensor $P_{\mu\nu}$ reduces to the Minkowski metric tensor $\e_{\mu\nu}=(1,-1,-1,-1)$ in the zero acceleration limit.  For example, the non-vanishing Poincar\'e metric tensors $P_{\mu\nu}$ for a class of frames $F(w,x,y,z)$ with constant linear acceleration $\a_o$ take the form: $P_{\mu\nu}=(W^2, -1,-1,-1),$ where $W^2=\g^2(\g_o^{-2} + \a_{o} x), \ \g=1/\sqrt{1-\b^2}, \ \g_o = 1/\sqrt{1-\b^2_o}, \ \b = \a_o w + \b_o.$\cite{12}  For a more general case such as frames with arbitrary linear accelerations, see ref. 4.} and the $T_4$ generators have the familiar representation $p_{\mu}=i\p_{\mu}$.  The $T_4$ generator differs from the generators of all other internal gauge groups, which have constant matrix representations. 

There are five independent gauge coupling constants:  $g_s$, $g_b$, $f$ and $f'$ are dimensionless, and $g$ has the dimension of length because the $ T_4$ generator $p_{\mu}=i\p /\p x^\mu$ has the dimension of  (1/length)  in natural units.     A basic difference between gravity and, say, electrodynamics is as follows:  The electron and the positron have the same attractive gravitational force characterized by one coupling constant $g$.  However, they have different electromagnetic forces (i.e., attractive and repulsive), which are characterized by two dimensionless coupling  constants $\pm {e}$.  This difference stems from the absence of an `$i$' in the $ T_4$ term involving $-ig \phi_{\mu}^{\nu}p_{\nu}= g\phi_{\mu}^{\nu}\p_{\nu}$ in (2).  This property also indicates that gravity described by $\phi_{\mu\nu}$ is the manifestation of the translational gauge symmetry in flat space-time. 

The gauge curvature associated with each individual group can be obtained by calculating the commutator of the covariant derivative defined in (2).  The commutator of the total-unified  gauge covariant derivative $\d_\mu$ defined in (2) is given by
\be
[\d_{\mu} ,\d_{\nu}]=  C_{\mu\nu\s} D^{\s} + ig_s {\bf G}^{a}_{\mu\nu} \frac{\ld^a }{2} +if W_{\mu\nu}^{ k} t^k  
+if' U_{\mu\nu} t_o  
+i\frac{g_{b}}{3} B_{\mu\nu}, 
\ee
%%%%%%%%%1%%%6%%%%C4%%%%%%%%16.29%%%3
where the $T_4$ gauge curvature  $C_{\mu\nu\s}$ is given by
\be
C_{\mu\nu\a}= J_{\mu\s}D^{\s} J_{\nu\a}-J_{\nu\s} D^{\s} J_{\mu\a}=\D_{\mu}J_{\nu\a}-\D_{\nu}J_{\mu\a}=-C_{\nu\mu\a},
 \ee 
 %%%%%%%%%%%%%%4
 $$\D_{\mu}=J_{\mu\nu}D^{\nu}, \ \ \ \ \ \   J_{\mu\nu}= P_{\mu\nu} + g\phi_{\mu\nu}, \ \ \ \ \ \   \phi_{\mu\nu} = \phi_{\nu\mu}=P_{\nu\ld}\phi_{\mu}^{\ld},$$
where $\D_{\mu}$ is the $T_4$ gauge covariant derivative in arbitrary coordinates with metric tensors $P_{\mu\nu}$ within flat space-time.   The modified color $SU_3$ curvature ${\bf G}^{a}_{\mu\nu }$  is given by
\be
{\bf G}^{a}_{ \mu\nu} =  J_{\mu}^{\s}D_{\s}G^{a}_{ \nu} - J_{\nu}^{\s}D_{\s}G^{a}_{ \mu} - g_{s}f^{abc}G^{b}_{ \mu}G^{c}_{ \nu}, 
\ee
%%%%%%%3%%%%%%7%%%%%%%%5
where $a,b,c=1,2,...8, $ and the structure constants $f_{abc}$ are completely anti-symmetric.\cite{5} 
The modified $SU(2) \times U(1)$  gauge curvatures $W_{\mu\nu i}$ and $U_{\mu\nu}$, in the presence of the gravitational gauge potential, are given by 
\be
W_{ \mu\nu}^{ i} =  J_{\mu}^{\s}D_{\s}W_{ \nu}^{ i} - J_{\nu}^{\s}D_{\s}W_{ \mu}^{ i} - f \epsilon^{ijk}W_{ \mu}^{ j}W_{ \nu}^{ k}, 
\ee
%%%%%%%3%%%%%%7%%%%%%%%30%%%%%%4%%6
\be
U_{\mu\nu} =  J_{\mu}^{\s}D_{\s}U_{ \nu} - J_{\nu}^{\s}D_{\s}U_{ \mu}, \ \ \ \ \ \   J_{\mu}^{\s} = \d_{\mu}^{\s} + g\phi_{\mu}^{\s},
\ee
%%%%%%%%%31%%%%%7
where $i,j,k=1,2,3,$ for the SU(2) group.
The modified $U_{1b}$ gauge curvature $B^{\mu\nu}$ associated with the baryon charge is found to be
\be
B_{\mu\nu} =  J_{\mu}^{\s}D_{\s}B_{ \nu} - J_{\nu}^{\s}D_{\s}B_{ \mu},
\ee
%%%%%%%%%%%32%%%%%%%8
in the presence of gravity.

Let us consider the physical implications of the total-unified model due to the presence of gravity.  For simplicity, we use inertial frames with the metric tensor $P_{\mu\nu} =\e_{\mu\nu}$ and $D_{\mu}=\p_{\mu}$.  We first note that all these modified gauge curvatures (5)-(8) reduced to the usual gauge curvatures in quantum chromodynamics, electroweak theory and $U_{1b}$ gauge theory in the limit
\be
g \to 0, \ \ \ \  or  \ \ \ \ \  J^\s_{\mu} \to \d^\s_\mu.
\ee
%%%%%%%%9
In particular, since $SU_2$ and $ U_1$ gauge transformations  are the same as those in the Weinberg-Salam theory,\cite{5}    the local gauge transformations for $W_{\mu}^{i}(x) $ and $W_{\mu\nu}^{ i}(x)$ are given by
\be
W_{\mu}^{' i}(x) =W_{\mu}^{i} (x)+ \frac{1}{f} \p_\mu \om^i (x)+C^{ijk} \ \om^{j}(x)W_{\mu}^{k}(x),
\ee
%%13%%%%%%10
\be 
W_{\mu\nu}^{' i} (x) =W_{\mu\nu}^{i} (x) + C^{ijk} \ \om^j (x)W_{\mu\nu}^{ k}(x), \ \ \ \  for \ \ \ \    g=0
\ee
%%%%%14%%%%%%%%%%11
where 
\be
W_{ \mu\nu}^{ i} =  \p_{\mu}W_{ \nu}^{ i} - \p_{\nu}W_{ \mu}^{ i} - f \epsilon^{ijk}W_{ \mu}^{ j}W_{ \nu}^{ k}
\ee
%%%%12
 for $SU_2$ gauge curvature.\cite{5}  
 
\section{Very small  violations  of  $U_1$, $SU_2$ and $SU_3$ gauge symmetries  due to gravity}
 
 First, we note that, under the $U_{1b}$ gauge transformation, we have
$$
 B'_\mu(x)=B_{\mu}(x)+ \p_\mu \Ld (x), \ \ \ \ \     
$$
\be
B'_{\mu\nu}(x) = B_{\mu\nu}(x),   \ \ \ \  \p_\ld B'_{\mu\nu}(x) = \p_\ld B_{\mu\nu}(x),   \ \ \ \  for   \ \ \  g=0.
\ee
%%%%%13
Thus, in the absence of gravity, i.e., $g=0$, both $B_{\mu\nu}(x)$ and $\p_\ld B_{\mu\nu}(x)$ are gauge invariant.  Both the quadratic forms $ L_1 =-(1/4)B_{\mu\nu}(x)B^{\mu\nu}(x)$ and $L_2=(-L_s^2 /4)\p_\ld B_{\mu\nu}(x) \p^\ld B^{\mu\nu}(x)$ can be used as the invariant Lagrangian of  the baryonic gauge field.  The Lagrangian $L_1$ was originally discussed by Lee and Yang to understand conservation of the baryon number.\cite{6}  Such a baryonic field is identical to the electromagnetic field, except that the original Lee-Yang force for baryons is much weaker than the gravitational force, so that it has no other observable effects in physics.
We shall assume the modified gauge invariant Lagrangian $L_2$ for the baryonic gauge field because it leads to a linear static potential for the baryonic gauge field and it can have cosmological implication on the accelerated expansion of the universe involving baryon matter.

However,  in the presence of gravity, the gauge curvature  $ B_{\mu\nu}$ is modified to have the form (8) when $g \ne 0$.  Under the $U_{1b}$ gauge transformation, instead of (13), we obtain
\be
 B'_{\mu\nu} =  B_{\mu\nu}+ J_\mu^\s \p_\s \p_\nu \Ld(x) - J_\nu^\s \p_\s \p_\mu \Ld(x)\ne B_{\mu\nu}.
 \ee
 %%%%%%%%%14
Thus, the modified gauge curvature $B_{\mu\nu}$ in (8) has a very small violation of $U_{1b}$ gauge invariance due to gravity.  Similar non-gauge invariance happens to the $SU_2$  and $SU_3$ gauge curvature.  As a result, there are very small violations of the internal gauge symmetries in all gauge curvatures (5)-(8)  due to gravity. 

Such an extremely small violation of internal gauge symmetries due to gravity is intimately related to the property that all gauge fields and generators associated with the groups $T_4,\ SU_3, \ SU_2$ and $U_1$ are on equal footing in the total gauge covariant derivative ${\d}_\mu$ in (2), just like that in the $SU_2\times U_1$ electroweak theory.  This `equal footing' of all gauge fields in (2) leads to the simplest coupling between gauge fields and should be considered as a basic postulate of total-unification of external and internal gauge symmetries in the model.

Gauge transformations of internal groups $SU_2$ and $U_1$, etc are well-known.\cite{5}  Some of them are discussed in equations (10)-(13). However, the $ T_4$ gauge transformations are more involved because the (infinitesimal) local and arbitrary space-time translations, 
\be
x'^{\mu} = x^{\mu}+\Ld^{\mu}(x),
\ee
%%%%%%%%%%%%%%21%%13%%%19****15
are, simultaneous, also the general coordinate transformations with arbitrary infinitesimal function $\Ld^{\mu}(x)$.  Thus, the $ T_4$ gauge transformations for, say, vector and tensor fields\footnote{For generality, in the rest of this section, we shall consider arbitrary coordinates associated with the Poincar\'e metric tensor $P_{\mu\nu}$ in a general frame of reference  within the flat space-time} are formally the same as the Lie variation of tensors.
Under the infinitesimal $T_4$ gauge transformations in flat space-time,\cite{9} we have
$$
S^{\$}(x)= S(x) -\L^{\s}(x) \p_{\s} S(x)
$$
\be
 \Phi^{\$}_{\mu}(x)= \Phi_{\mu}(x) -\L^{\s}(x) \p_{\s} \Phi_{\mu}(x) - \phi_{\s}(x)\p_{\mu} \L^{\s}(x), 
\ee
%%%%16%%20*****16
$$
 \phi^{\$}_{\mu\nu}(x)= \phi_{\mu\nu}(x) -\L^{\s}(x) \p_{\s} \phi_{\mu\nu}(x) - \phi_{\mu\s}(x)\p_{\nu} 
\L^{\s}(x) - \phi_{\s\nu}(x)\p_{\mu} \L^{\s}(x), 
$$
%%%%29%%%17%%%%21*******
and so on, where $\Ld_{\mu}(x)$ denotes infinitesimal and arbitrary vector functions.  Furthermore, we
 assume that the tensor gauge field $\phi_{\mu}^{\nu}(x)$ is an electrically neutral field and an iso-scalar, so that the $ T_4$ gauge potential $\phi_{\mu}^{\nu}$  does not transform under  $SU_3, \ SU_2$ and $ U_1$ gauge transformations. 
 
 To see the exact invariance of the total-unified action functional  $S_{tu}=\int L_{tu}\sqrt{-P}d^{4}x$, where $P=det P_{\mu\nu}$, in a general reference frame (inertial and non-inertial) with arbitrary coordinates associated with Poincar\'e metric tensor $P_{\mu\nu}$,\cite{9,12} we first note that the total-unified Lagrangian $L_{tu}$ is a scalar function by construction with the gauge curvatures given by equations (4)-(8) in arbitrary coordinates. Thus it transforms the same as a scalar $S(x)$ in (16).  One can show that $\sqrt{-P}$ transforms as follows,
 \be
 \sqrt{-P^{\$}} =[(1-\Ld^\s\p_\s)\sqrt{-P}](1-\p_{\a}\Ld^{\a}),
 \ee
 %17
 where we have used the last relation in (16).\cite{1}  We have
 \be
 S_{tu}^{\$}= S_{tu} -\int[\p_{\ld}(\Ld^{\ld} L_{tu}\sqrt{-P})d^4 x = S_{tu},
 \ee
 %%%%%18
where the divergent term inside the four-dimensional volume integral does not contribute because it can be transformed into the integral of a vector over a hypersurface on the boundaries of the volume of integral where fields and their variation vanish.  Therefore, the total-unified action $S_{tu}$ is exactly invariant under the infinitesimal $T_4$ gauge transformations (16).  In contrast, all other internal gauge symmetries have a very small non-invariance in their gauge curvatures due to the presence of gravity in the total-unified model, as we have discussed previously.

\section{Feynman-Dyson rules in the strong sector (chromogrvity)}

The rules for Feynman diagrams without involving gravitons in the usual electroweak theory and quantum chromodynamics (QCD) are well-known.\cite{5}  The rules in pure Yang-Mills gravity were obtained and discussed.\cite{13}

In the following discussions, we shall consider the total-unified model in inertial frames for simplicity.  Within the generalized Yang-Mills framework in inertial frames with $P_{\mu\nu}=\e_{\mu\nu}$, all covariant derivatives $D_\mu$ in (2)-(8) are replaced by $\p_{\mu}$.  As usual, the total-unified Lagrangian is assumed to be quadratic in gauge curvatures,
$$
L_{tu}= L_{\phi} + L_G  -\frac{1}{4}(W^{\mu\nu k}W_{\mu\nu}^k + U^{\mu\nu}U_{\mu\nu}),
$$
%%%%%%%%%33%%%%15%%19
\be
 - \frac{1}{4}L_s^2 \p_{\ld}B_{\mu\nu} \p^{\ld}B^{\mu\nu} +L_{matter} + L_{H},
\ee
\be
L_{\phi}= \frac{1}{4g^2}\left (C_{\mu\nu\a}C^{\mu\nu\a}- 2C_{\mu\a}^{ \ \ \  \a}C^{\mu\b}_{ \ \ \  \b} \right), \ \ \  \ee
 %%%%9%%%%%%%%%6%%%16
 \be
L_{G}= -\frac{1}{4}{\bf G}^{a\mu\nu}{\bf G}^{a}_{\mu\nu}, 
\ee
%%%%%%%%%17
%\be
%  L_{q}=+ \overline{q}(i\g^{\mu}{\bf d}_{\mu}- M)q,     
%\ee 
%%%%%%%10%%%%%%%%%%%%7%%%%18%%%%%%%0000000
%$$
%      \{\g_\mu, \g_\nu\} = 2 \eta_{\mu\nu},    
%$$
where  the Lagrangian $L_H$ in (18) involves Higgs scalar doublets and their couplings.  The Lagrangian $L_{matter}$ will be discussed later in section 6.  It  contains spinor quarks with three colors and six flavors and their interactions with gauge fields.\cite{5,1,2}  

We now use the total-unified Lagrangians (19) to write down the Feynman-Dyson rules for the interactions of gravitons, gluons and quarks. Although these are conventionally referred to as the Feynman rules,\cite{14} we choose to refer to them as the Feynman-Dyson rules in this total unified model because the intuitive Feynman rules for calculating the invariant amplitude (a Lorentz scalar) are inadequate for higher-orders.  It was Dyson's rigorous derivation and elaboration of the rules that allowed their application to complicated and higher-order interactions in quantum Yang-Mills gravity and in Kinoshita's higher-order corrections to the magnetic moment of the electron in quantum electrodynamics.\cite{15,16}

Let us consider the general propagator for the graviton corresponding to the general 
gauge-fixing terms 
\be
L_{\xi\z}=\frac{\xi}{2g^{2}}\left[\p^\mu J_{\mu\a} - 
\frac{\z}{2} \p_\a J^\ld_\ld\right]\left[\p^\nu J_{\nu\b} - 
\frac{\z}{2} \p_\b J^\ld_\ld\right]\e^{\a\b}.
\ee
%%%%%%%%%%%%4%%%%22%%%21%%%22
From (19) and (22) with $\zeta=1$ and $\xi=2$ , we obtain     the following simple propagator for the graviton $G_{\a\b\rh\s}$,
\be
G_{\a\b\rh\s}=\frac{-i}{2k^2}\left[\frac{}{}(\e_{\a\b}\e_{\rh\s}- \e_{\rh\a}\e_{\s\b}-\e_{\rh\b}\e_{\s\a})\right],
\ee
%%%%%%%%13%%%%17%%%%%%%23%%%%%24%%%%23
where the usual prescription of $i\ep$ for the Feynman propagator (23) is 
understood.   In this paper, it should always be understood as being done, even when the $i\ep$ is not explicitly included in the propagators.   The propagator (23) has been discussed by DeWitt and the gauge condition specified by (22) with $\zeta=1$ and $\xi=2$  may be called the DeWitt gauge.\cite{17} The result (23) is consistent with that obtained in  previous 
works by Leibbrandt, Capper and others in general relativity.\cite{18,19,20} As an aside, the overall factor of (1/2) in the graviton propagators (23) is necessary for Yang-Mills gravity to be consistent with its $T_4$ gauge identity---a generalized Ward-Takahasi identity for the Abelian group $T_4$ with ghosts and to be consistent with the Slavnov-Taylor identity.\cite{21}

We now consider the Feynman-Dyson rules for graviton interactions corresponding  to
$iL_{G}$ in (21).  The gluon and graviton momenta in the Feynman-Dyson rules are in-coming to the vertices.
Suppose  the 3-vertex for gluon-gluon-graviton is
denoted by ($G^{\mu a}(p)G^{\nu b}(q)\phi^{\a\b}(k)$).  The gluon-gluon-graviton 3-vertex is
\be
ig\de^{ab}  \left(\frac{}{}p^{\a}q^{\b}\e^{\mu\nu}
+p^{\b}q^{\a}\e^{\mu\nu}- p^{\nu}q^{\b}\e^{\mu\a}
- p^{\b}q^{\mu}\e^{\a\nu}\right)_{(\a\b)}.
\ee
%%%%%%%%%%%%%15%%%%%%%%19%%%%%%%%%%%%25%%%%%%%26%%%24
We use the symbol ${(\a\b)}$ to denote that a symmetrization is to be performed on the  index 
pair $(\a\b)$. 

The 4-vertex for gluon-graviton coupling,  
($ G^{\mu a}(p) G^{\nu b}(q)G^{\s c}(k)  \phi^{\a\b}$), is given by
$$
{ig g_s}f^{abc} \left(\frac{}{}p^{\b}\e^{\mu\s}\e^{\nu\a}- \
p^{\b}\e^{\a\s}\e^{\mu\nu}- q^{\b}\e^{\mu\a}\e^{\nu\s}\right.
$$
\be
\left. + \ q^{\b}\e^{\a\s}\e^{\mu\nu}+ k^{\b}\e^{\mu\a}\e^{\nu\s}- \ k^{\b}\e^{\mu\s}\e^{\a\nu}\right)_{(\a\b)} 
\ee
%%%%%%%%%%16%%%%%%%%%%20%%%26%%27%%%25
where $f^{abc}$ is the completely anti-symmetric structure constant of the color $SU_3$ group.
There is another 4-vertex for 2-gluon and 2-graviton coupling, denoted by
 $(G^{\mu a}(p)G^{\nu b}(q)\phi^{\a\b}\phi^{\g\de})$, which  is given by
$$
ig^2 \de^{ab} \left(\frac{}{}p^{\b}q^{\de}\e^{\mu\nu}\e^{\a\g}
+p^{\de}q^{\b}\e^{\mu\nu}\e^{\a\g}\right.
$$
\be
\left.- p^{\b}q^{\de}\e^{\mu\g}\e^{\nu \a}
- p^{\de}q^{\b}\e^{\a\mu}\e^{\nu\g}\right)_{(\a\b)(\g\de)},
\ee
%%%%%%%%%26
where both  index pair ${(\a\b)}$ and ${(\g\de)}$ are to be symmetrized.

%The quark propagator has the usual form
%\be
%\frac{i}{\g^{\mu} p_{\mu} - m} \de_{jj'} \de_{ff'},
%\ee
%%%27%%%28
%where $j$ and $f$ are respectively color and flavor indices. 
%(27)????

\section{New photon-graviton couplings}

In this section, let us consider the Feynman-Dyson rules for new graviton-photon interactions in  the QED sector implied by the matter Lagrangian $L_{matter}$ in (15).  The Lagrangian $L_{\g}$ with new graviton-photon coupling can be derived from the gauge boson terms involving $W^{\mu\nu a}$ and $U^{\mu\nu}$ in (19) and is given by\cite{1,5}
\be
L_{\g}=\frac{-1}{4}F_{\mu\nu}F^{\mu\nu},  
\ee
%%%%%%%%%%%%%30%28****27
$$
F_{\mu\nu}=J_{\mu\s}\p^{\s} A_\nu - J_{\nu\s}\p^\s A_\mu,\ \ \ \ \  J_{\mu\s} = \e_{\mu\s} + g\phi_{\mu\s},
$$
where $A_\mu$ is the electromagnetic potential and $F_{\mu\nu}$ is the modified electromagnetic field strength in the presence of gravity.  The 3-vertex for the photon-graviton coupling is
denoted by ($A^{\mu}(p)A^{\nu}(q)\phi^{\a\b}(k)$) and given by  
\be
ig  \left(\frac{}{}p^{\a}q^{\b} \e^{\mu\nu}
+p^{\b}q^{\a} \e^{\mu\nu}- p^{\nu}q^{\b} \e^{\mu\a}
- p^{\b}q^{\mu} \e^{\a\nu}\right)_{(\a\b)}.
\ee
%%%%%%%%%%%%%15%%%%%%%%19%%%%%%%%%%%%25%%%%%%%26%%%31****28
We use the symbol ${(\a\b)}$ to denote that a symmetrization is to be performed on index 
pair $(\a\b)$.  All photon and graviton momenta are incoming to the vertex.

The 4-vertex for the photon-graviton coupling,  
($ A^{\mu}(p) A^{\nu}(q)A^{\s}(k)  \phi^{\a\b}$), is given by
$$
 ig^2 \left(\frac{}{}p^{\b}q^{\de} \e^{\mu\nu}\e^{\a\g}
+p^{\de}q^{\b} \e^{\mu\nu}\e^{\a\g}\right.
$$
\be
\left.- p^{\b}q^{\de} \e^{\mu\g}\e{\nu \a}
+ p^{\de}q^{\b} \e^{\a\mu}\e^{\nu\g}\right)_{(\a\b)(\g\de)},
\ee
%%%%%%%%%27%32****29
where both  index pair ${(\a\b)}$ and ${(\g\de)}$ are to be symmetrized.

\section{The matter Lagrangian for quarks, leptons with  gravitons}

Let us now consider the interactions of gravitons with quarks and leptons.  As usual, we denotes left-handed and right-handed fermions (weak interaction or gauge eigenstates) as follows:  
\bigskip

%%%%%%%%%%%%%%%
\bed
%\math
  \ \ \ \ \  L: \ \ \ \  \left(\begin{array} {cc}
\nu_e  \\
e\\
\end{array}\right)_L  \ \ 
%\math
\left (\begin{array} {cc}
\nu_{\mu} \\
\mu\\
\end{array}\right)_L \ \  
%\math 
\left (\begin{array} {cc}
\nu_{\tau}\\
\tau\\
\end{array}\right)_L  \ \ ..... 
\eed

%\bigskip

$$ \ \      \overline{R}: \ \ \ \ \ \ \      \overline{e_R} \ \ \ \ \ \ \ \ \ \ \    \overline{\mu_R} \ \ \ \ \ \ \ \ \ \ \ \    \overline{\tau_R} \ \ \ .....$$

\bed
%\math 
 \ \ \ \   Q_L: \ \ \   \left(\begin{array} {cc}
u  \\
d\\
\end{array}\right)_L  \ \ \ 
%\math
\left (\begin{array} {cc}
c \\
s\\
\end{array}\right)_L \ \ \  
%\math 
\left (\begin{array} {cc}
t\\
b\\
\end{array}\right)_L  \ \ \  .....
%\bigskip
\eed
%\bigskip

$$    \overline{Q_R}:  \ \ \ \ \ \     \overline{u_R}  \ \ \ \ \ \ \ \ \ \  \overline{c_R} \ \ \ \ \ \ \ \ \ \ \  \overline{t_R} \ \ \ ....$$
%\bigskip

$$ \     \overline{Q'_R}: \ \ \ \ \ \    \overline{d_R} \ \ \ \ \ \ \ \ \ \  \overline{s_R} \ \ \ \ \ \ \ \ \ \ \  \   \overline{b_R}
 \ \ \ ....$$
\bigskip

\noindent
By definition, the $SU_3$ generator $ g_s \ld_3/2$ has eigenvalues: $+g_s/2$ for the red quark doublet and the green antiquark singlets; $-g_s/2$ for the green quark doublet and the red antiquark singlets; and zero for all other left-handed fermions.\cite{22,23}

In this total-unified model in inertial frames, the action $S_{tot}$ is assumed to be
\be
S_{tu} = \int d^4 x \ L_{tu}, 
\ee
%%35%%%%************34*****32****30
where $L_{tu}$ is given by (19), in which the matter Lagrangian $L_{matter}$ involving quarks and leptons is given by
\be
L_{matter}=\overline{L} \ i\g^\mu\de_\mu L +\overline{R} \ i\g^\mu\de_\mu R+\overline{Q_L} \ i\g^\mu\de_\mu Q_L
\ee
%38*****33*****31
$$
+\overline{Q_R}  \ i\g^\mu\de_\mu Q_R+\overline{Q'_R} \  i\g^\mu\de_\mu Q'_R,
$$
where we have suppressed the family index.  Let us concentrate on the coupling of gravitons and fermions in the total-unified model.  The total gauge covariant derivatives $\de_\mu$ applied to fermion fields give the results,
$$
(\de_\mu Q_L)_{fm}=\left[\frac{}{}\d_{fh}\d_{mj}(\p_\mu + g\phi_\mu^{\nu }\p_\nu +if' U_\mu t^o + ig_b B_\mu)\right.
$$
\be
\left.+\d_{fh}(ifW^k_{\mu} t^k)_{mj} +( ig_s G^a_\mu \frac{\ld^a}{2})_{fh}\d_{mj}  \right](Q_L)_{hj},
\ee
%%%%39*****34***32
$$
(\de_\mu Q_R)_{f}=\left[\frac{}{}\d_{fh}(\p_\mu + g\phi_\mu^{\nu }\p_\nu  +if' U_\mu t^o + ig_b B_\mu) \right.
$$
\be
\left.+( ig_s G^a_\mu \frac{\ld^a}{2})_{fh}\right](Q_R)_h,
\ee
%%40****35***33
$$
(\de_\mu Q'_R)_{f}=\left[\frac{}{}\d_{fh}(\p_\mu + g\phi_\mu^{\nu }\p_\nu  +if' U_\mu t^o + ig_b B_\mu) \right.
$$
\be
\left.+( ig_s G^a_\mu \frac{\ld^a}{2})_{fh}\right](Q'_R)_h,
\ee
%%%%%41***36***34
$$
(\de_\mu L)_m=\left[\frac{}{}\d_{mj}(\p_\mu + g\phi_\mu^{\nu }\p_\nu  +if' U_\mu t^o + ig_b B_\mu)  \right.
$$
\be
\left.+(ifW^k_{\mu} t^k)_{mj}\right](L)_{j},
\ee
%%%%%42******37***35
\be
\de_\mu R=\left(\p_\mu + g\phi_\mu^{\nu }\p_\nu +if' U_\mu t^o + ig_b B_\mu\right)R,
\ee
%%%%%43****38***36
%%%%%%%
where the eight $[{SU_3}]_{color}$  generators are represented by $\ld^a/2$, a=1,2,...8, (i.e., the Gell-Mann matrices) with the normalization, $Tr(\ld^a  \ld^b)=2 \de^{ab}$; $t^k$ are the $SU_2$ generators in general (while their fundamental representation is denoted by $\tau^k/2$); and $t^o$ is the hypercharge generator of the $U_1$ group.\cite{5, 23} 

In the absence of gravity and baryon field (i.e., $g=g_b=0$), the matter Lagrangian (31) reduces to the corresponding results in Weinberg-Salam theory and quantum chromodynamics.\cite{23}  In general, it is lengthy to write down the results in terms of physical leptons and quarks.  Let us consider the Feynman-Dyson rules for the couplings of graviton-leptons and graviton-quarks in the matter Lagrangian (31).
For example, the Feynman-Dyson rules for the quark-graviton 3-vertex  
($\overline{q}(k) q(p) \phi_{\mu\nu}$), is given by 
\be
 \ \frac{i}{2} g\g_{\mu}(p_{\nu} + k_{\nu})_{(\mu\nu)}, 
 \ee
 %%%%%%%%%%%27***37
where $q(p)$ may be considered as an annihilation operator of a 
quark with momentum $p_{\nu}$, and $\overline{q}(k)$ may be a creation 
operator of a quark with the momentum $k_{\nu}$.

The electron-graviton 3-vertex with 
($\overline{e}(q) e(p) \phi_{\mu\nu}(k)$) has the same form as (37),
\be
 \ \frac{i}{2} g\g_{\mu}(p_{\nu} + q_{\nu})_{(\mu\nu)},
\ee
%%%%%%%%28%%29%%%34***38
where the electron in the Lagrangian has been symmetrized.\cite{13}  Note that $e(p)$ may be considered as an annihilation operator of an electron  with momentum $p_{\nu}$, and $\overline{e}(q)$ is considered as a creation 
operator of an electron with the momentum $q_{\nu}$.

We may remark that the total Lagrangian (19) is gauge invariant under the $SU_3\times SU_2 \times U_1 \times U_b1$ only in the limit $g\to 0$, i.e., the absence of gravitational  interaction.  In general, the total action (30) is an exactly gauge invariance under local space-time translational gauge group.  All internal gauge symmetries such as $SU_3, SU_2$ and $U_1$ are only approximate in the presence of  gravitational interaction.  The physical effects of these violations are in general too small to be detected in laboratory, unless a macroscopic object such as the sun or a galaxy is involved, as we shall see in the next sections.

\section{Experimental tests of Lee-Yang force for accelerated expansion of the universe  }
%\bigskip

  An interesting question is:  Can the accelerated expansion of the universe be understood within the total-unified model?  This great enigma challenges the power of Yang-Mills idea of gauge symmetry.  
  
  A linear potential and its effect on the accelerated expansion of the universe have been discussed.\cite{7}  Here we summarize the result in the context of total-unified model. 
First, we observe that the result (13) implies that the gauge invariance of $\p_\ld B_{\mu\nu}$ enable us to construct an alternative gauge invariant Lagrangian $L_{AE}$ for  involving up and 
down quarks and baryonic gauge field $B_{\mu}$
\be
L_{AE}= -\frac{L_s^2}{4} \p_\ld B_{\mu\nu} \p^\ld B^{\mu\nu} 
 + L_{ud},   
 \ee
 %%%%%%%%%% 16.1%%%22%%%43***39
$$
L_{ud}=i\overline{u}_{a} \g_\mu 
(\p^\mu-\frac{ig_{b}}{3}B_{\mu})u_{a} - 
m_{u}\overline{u}_{a}u_{a} + (u \rightarrow d),
$$
%%%%%%2%%23*****???
where the color index $a$ is summed from 1 to 3.   

For accelerated expansion, let us ignore the gravitational fields for simplicity and concentrate on the baryonic gauge fields and the associated Lee-Yang force.  The new gauge-invariant field equation derived from (39) is a  fourth-order partial differential equation,
$
 \p^2\p_\nu B^{\nu\mu} - g'_b J_{q}^\mu = 0, $
%%%%%%%%%%%%12%%%%%%%%%%10%%%%35%%%%24*********
where  $g'_b = g_b/(3 L_s^2)$, and   
 the source of the gauge field is 
$J_{q}^{\mu} =  \overline{u}_n \g^{\mu}u_n  +
  \overline{d}_n \g^{\mu}d_n.$
For the static case, the potential field $B_{0}$ satisfies the fourth order 
differential equation
\be
\bigtriangledown^2 \bigtriangledown^2 B^0 
= -\frac{g_{b}}{3L_{s}^{2}}J^{0} \equiv \rho_{B}.  
\ee
%%%%%%%%33%%%%%37%%%%4%%%25%%%%44***40

Suppose we impose a `Coulomb-like gauge' $\p_k  B^k = 0$, the static exterior potential
satisfies the equation $\bigtriangledown^2 \bigtriangledown^2 B^0 = 
0$. The solution for such a potential is not unique, one has 
$B^{o}=A'/r$,   or  $ B'r$   or  $ C'r^{2}.$ 
  It turns out that only the solution  
$B'r$ can be understood as due to the exchange of virtual quantum described by 
the fourth order field equation.  The reason is as follows:

From the viewpoint of quantum field theory,
the fermion (baryon) source $\rho_{B}$ in (40) is represented by the usual 
delta-function because these fermions satisfy the Dirac 
equations.  Based on this property, 
the solution for the potential $B_0$ should be proportional to $r$
rather than $r^2$ or $1/r$.   This result can be seen explicitly by
substituting $\rho_{B}({\bf r})=g'_{b}\delta^{3}({\bf r})$ in (40), 
where $g'=g_{b}/(3L_{s}^{2})$ for quarks (and $g'=g_{b}/(L_{s}^{2})$ for 
protons and neutrons).   We obtain a linear potential in the static limit,
\be
   B^{0}({\bf r}) =g'_{b}\int_{-\infty}^{\infty}{\frac{1}{({\bf k}^2)^2} e^{i{\bf k}\cdot{\bf r}} d^3
k} = - \frac{g'_{b}}{8\pi}|{\bf r}|,   
\ee
%%%%%12%%%%%%%%%%%%39555516.6%%%%%27%%%%28%%%45***41
which can be understood as a generalized function.\cite{24}  
This linear potential (41) could produce a  constant repulsive Lee-Yang force between baryonic galaxies. 

The equation of motion of a freely moving test particle in the Newtonian limit is changed to the following form\cite{7}
\be
\frac{d^2 {\bf r}}{dt^2} = {\bf g} + {\bf g}_{LY},
\ee
%%16.7%%%%%%%28%%%42%%%%%%%46***42
where ${\bf g}$ is the gravitational acceleration produced by the distribution of ordinary matter,\cite{25} while the constant  acceleration ${\bf g}_{LY}$ is due to the modified Lee-Yang's baryon gauge field.\cite{7}  This constant repulsive force between baryon matters in the universe is the characteristic feature of the modified cosmic Lee-Yang force.    If one uses Einstein's field equation with a cosmological constant one will have the result that the acceleration of expansion depends linearly on the distance between two galaxies.\cite{25,7}  Thus, a crucial question is  whether the acceleration in the expansion of the universe is a constant or dependent of distance.  Since we have accelerated Wu transformations of space-time, which gives new Doppler shift for radiation source with constant acceleration,\cite{12} it is hoped that experimental test of (42) can be carried in the near future.

Because of the extreme weakness of the coupling strength of the baryonic gauge field $B_\mu$, its existence cannot be detected in the solar system and in our galaxy, it may be fitting to call it a ``dark field."  However,  the sources of such a dark field are the ubiquitous baryonic matter rather   than some mysterious `dark energy'  in the universe.  This dark field model suggests that the accelerated cosmic expansion could be understood in terms of a modified Lee-Yang force produced by a baryonic gauge field.  Thus, the Lagrangian (39) for the ``dark field" $B_\mu$ has a field-theoretic interpretation and can be included in the total unified model.  

\section{Experiments for geometric-optics limits of wave equations for quarks and gauge bosons}

Next, let us consider the relation between the classical 
equation of motion and the massive fermion wave equation.  The fermion 
wave equation  can be derived from the Lagrangian (19) in inertial frames,\cite{9}
\be
i\g_\mu J^{\mu\nu}\p_{\nu} \psi - m \psi + \frac{i}{2} \g_{\mu}[\p 
_{\nu}J^{\mu\nu}] \psi = 0,    
\ee
%29%%%%%%%46%%%%47888*****43
where the fermion fields have been symmetrized.
Using the expression for the field
$\psi= \psi_{o}exp(iS)$,\cite{26} we can derive the equation
for the motion of a classical objects in the presence of the 
gravitational tensor field $\phi^{\mu\nu}$ in flat space-time,  
 \be
G^{\mu\nu}(\p_{\mu}S)(\p_{\nu}S) - m^{2} = 0,  \ \ \ \ \ \ \  G^{\mu\nu} = P_{\a\b} J^{\a\mu} J^{\b\nu}. 
\ee
%%%%8.40%%%%%%%%%%32%%47%%%%%%%48***44
It is formally the same as the corresponding Hamilton-Jacobi equation in general relativity.\cite{27}  Within the framework of Yang-Mills gravity in flat space-time, we shall called (44) the Einstein-Grossmann equation of motion for classical objects.\footnote{Marcel Grossmann was associated with Einstein in elucidating the mathematical foundation of general relativity.}  It appears that the classical equation of motion must have this particular form, (44), in order to describe accurately the motion of big objects in the presence of gravity, no matter whether the theory is based on curved or flat space-time.

In Yang-Mills gravity, the fundamental equation of geometrical
optics can be derived.    After some calculations, we find a new eikonal equation, which involves an effective metric tensor $G_L^{\mu\nu}$ for a light ray,
\be
G_L^{\mu\nu} \p_\mu \Psi \p_\nu \Psi  = 0,
\ee
%%%%%%%%3%%%6%%%%%%7%%%8.36$$$$$$$35%%%%%%%40%%%48%%%%%%49***45
\be
G_L^{\mu\nu}  = G^{\mu\nu} -\frac{g}{4}\phi_{\ld}^\mu J^{\ld\nu}
=\e_{\a\b}(\e^{\mu\a} + g \phi^{\mu\a})(\e^{\nu\b} + \frac{3g}{4} \phi^{\nu\b}),
\ee%%%%%%%%%%41
%%%%%%%%%%%%%%4%%%%%%%7%%%%%%%%%%%%%%%49%%%%%%50****46
in the geometric-optics (or classical) limit.    

Equations (44) for fermions and (45) for gauge bosons are not exactly the same due to their different gauge invariant interactions and different properties of fields. 

If the smaller term $- (g/4)\phi_{\ld}^\mu J^{\ld\nu}$ in the effective metric tensors ${G_L}^{\mu\nu}$ for the eikonal equations (46) is neglected, then both fermions and bosons have the same effective Riemannian metric tensors in the geometric-optics limit.
In view of the good agreement between the effective metric tensor $G^{\mu\nu}$ in (44) and the perihelion shift,\cite{9} the small additional term in ${G_L}^{\mu\nu}$ given by (46)  suggests a possible experimental test of Yang-Mills gravity.

Let us estimate the contribution of the smaller term $(g/4)\phi_\ld^\mu J^{\ld\nu}$ in (46) to the bending of a light ray by the sun.  In the presence of gravity, the electromagnetic wave equation with the usual gauge condition $\p_\mu A^\mu = 0$, leads to the eikonal equation for a light ray (45) 
in the short-wavelength (or geometric-optics) limit.   
Following   the usual procedure, Yang-Mills gravity predicts a smaller angle for the bending of a light ray,  
\be
\D \phi_{light} = \frac{7Gm}{ 2R_o}\approx 1.53'',
\ee
%%%%%%%%%%%%%%%%%7%%%%%%%%%%%36%%%%%%%42%%%%43%%51****47
where $R_o$ is the distance from sun's center.\cite{28,4}  The result (47) is smaller  than the usual value  $1.75''$ in Einstein gravity by about 12\%.  So far, the experimental accuracy  for the bending of a light ray (with optical frequencies) by the sun is no better than $10\% - 20\%$.\cite{29}

If the experimental accuracy can be further  improved to a few percent, the difference between Yang-Mills gravity and Einstein gravity can be tested.  The significance of this experiment cannot be over emphasized because it will indicate whether the Yang-Mills gravity with space-time translational gauge group dictates the gravitational interaction. 

We may remark that previous radar echo experiments are not suitable to test Yang-Mills gravity with the new eikonal equation (45).  The eikonal equation has the simple effective metric tensor $G_L^{\mu\nu}$ if and only if one takes
the geometric-optics (or high-frequency) limit.   The radar echo experiments used the frequencies 7840 MHz and 2300 MHz,\cite{30}  which are roughly $10^9$ (1/sec) and are too small in comparison with the frequencies of visible light ray $\approx 10^{14}$ (1/sec.).  The eikonal equation (45) is expected to be valid for high-frequency waves such as visible light rays.  

\section{Discussions and remarks}

In connection with linear potentials, we note that quark confinement could also be understood in terms of a linear potential with attractive force at low energies.\cite{8}
 For a linear potential to produce  only attractive force between quarks and quarks, quarks and anti-quarks, and anti-quarks and anti-quarks, the fields must be tensor of the second rank or scalar field.  But a scalar field cannot be a gauge field.  This property  suggests that  the relevant gauge symmetry is likely to involve both internal and external gauge symmetries.\footnote{It appears that gauge fields associated with only internal gauge symmetry  will lead to both attractive and repulsive forces, just like the electromagnetic force.}  It would be very interesting if the potentials at  the smallest distance scale inside hadrons between confined quarks and at
the large cosmic scale between galaxies with accelerated 
expansion are both due to linear potentials associated with the 
fourth-order differential field equations in the low energy 
approximation.\cite{7,8}  If this turns out to be the case, then `particle cosmology' 
would be very interesting in the physics of the future.

We may remark that an important difference between Yang-Mills gravity and Einstein gravity is that
the simple `effective Riemann metric tensor' $G_L^{\mu\nu}$ given by (45) is only for a high frequency light ray, while the inherent space-time metric tensor in Einstein gravity can be applied to an electromagnetic wave with any frequency.  Thus, if the radar wave in the echo experiment can be replaced by electromagnetic waves with higher frequencies, say, laser light, then it could also test Yang-Mills gravity versus Einstein gravity.

The total-unified model is not renormalizable by naive power counting. It is not clear whether  the $T_4$ gauge symmetry could change the situation at higher orders. The usual argument based on the running coupling constant may not be air-tight, because the gravitational coupling has not been included.  Thus, the usual properties and conclusions in gauge field theories cannot be taken for granted in the total-unified model.  Furthermore, the most complicated couplings of the graviton with itself  and other particles are the 4-vertex, as shown in (25), (26), (29) and ref. 13, in the total-unified model. In contrast, in the (Feynman) quantization of Einstein gravity, the graviton can have N-vertex of self-coupling, where N  is an arbitrarily large number.

In the total-unified model, Yang-Mills gravity reveals  the flat space-time translational origin of the gravitational field  $\phi_{\mu\nu}$ and leads to an {\em effective curved space-time} for the motion of classical objects.  It is a logically and experimentally consistent gauge theory, which represents an alternative to Einstein gravity.   Furthermore, Yang-Mills gravity with flat space-time translational gauge symmetry paves the way to unify all interactions based on a generalized Yang-Mills framework. In this sense, the total-unified model  is a satisfactory framework that could accommodate all fundamental laws of physics in a unified manner and could be consistent with the accelerated expansion of the universe.  These properties could be considered as an advantage of Yang-Mills idea of gauge symmetry and, furthermore, the total-unified model supports the idea that gauge symmetries dictate interactions, as advocated by Utiyama, Yang and others.\cite{11}

\newpage  
\bibliographystyle{unsrt}

\begin{thebibliography}{99}
\bibitem{7}J. P. Hsu, Mod. Phys. Letts. A, {\bf 20}, 2855 (2005).  It is assumed that the observable portion of the universe is dominated by baryonic matter, which is consistent with observations.
\end{thebibliography}

\end{document}